\documentclass[fp,twocolumn,letter]{jpsj3}
\usepackage{color}
\usepackage{bm}
\usepackage{mathrsfs}

\title{Notes on Ground-state Properties of Mixed Spin-1 and Spin-1/2 Lieb-Lattice Heisenberg Antiferromagnets}
\author{Yuhei Hirose$^{1}$, Shoma Miura$^{1}$, Chitoshi Yasuda$^{2}$, and Yoshiyuki Fukumoto$^{1}$}
\inst{$^{1}$Department of Physics, Faculty of Science and Technology, Tokyo University of Science, Noda, Chiba 278-8510, Japan\\
$^{2}$Department of Physics and Earth Science, Faculty of Science, University of the Ryukyus, Nishihara, Okinawa 903-0213, Japan}
\recdate{\today}
\abst{
Quantum Monte Carlo (QMC) simulations are performed to study ground-state properties of a mixed spin-1 and spin-1/2 Lieb-lattice Heisenberg antiferromagnet, in order to get further insight beyond the modified spin-wave (MSW) study reported in [J. Phys. Soc. Jpn {\bf86}, 014002 (2017)]. 
It is confirmed that the MSW results are in good agreement with the QMC results. In particular, the scaling relation found in the MSW study, which argues that sublattice spin reductions are inversely proportional to the sublattice sizes, is observed in our QMC simulation. We present a rigorous proof for spontaneous sublattice magnetizations induced by an infinitesimal uniform magnetic field. The calculation process in the MSW theory is reexamined to clarify the mathematical structure behind the scaling relation for sublattice long-range orders.
}
\begin{document}
\maketitle

The exploration of the quantum and frustration effects in spin systems is one of the most fascinating issues in condensed matter physics\cite{Heidelberg}. 
Recently, significant interest has been paid to spin-1/2 frustrated Heisenberg antiferromagnets on a diamond-like-decorated square lattice\cite{Hirose2016,Hirose2017} because of a possible relevance to the Rokhsar$-$Kivelson quantum dimer model\cite{QDM}.

For a diamond-like-decorated square lattice, if we define the interaction strength of four sides of a diamond unit as $J$ and that of the diagonal bond as $J'=\lambda J$, the ratio $\lambda$ determines the ground-state properties\cite{Hirose2017}.
For $\lambda_c<\lambda<2$, the ground state manifold consists of macroscopically degenerated tetramer-dimer states (see Fig.~3 (a) in Ref.~3), which is equivalent to the dimer covering of the square lattice.\cite{Hirose2016,Hirose2017,Morita}
For $\lambda<\lambda_c$, we obtain the ferrimagnetic ground state as shown in Fig. 3 (c) in Ref.~3, which is identical to a mixed spin system on the Lieb lattice in Fig.~1.
We define a mixed spin Hamiltonian on the Lieb lattice as follows,
\begin{align}
   H=\sum_{i\in \rm{A}}\sum_{j\in \rm{B}}J_{i,j}\bm{S}_{i}\cdot\bm{S}_{j},
\label{eq:1}
\end{align} 
where $J_{i,j}=1$ for nearest-neighbor pairs, otherwise $J_{i,j}=0$, $\bm{S}_{i}^2=S_{\rm{A}}(S_{\rm{A}}+1)$ for A-sublattice sites $i$ (closed circles in Fig.~1), and $\bm{S}_{j}^2=S_{\rm{B}}(S_{\rm{B}}+1)$ for B-sublattice sites $j$ (closed triangles).
The above-mentioned ferrimagnetic ground state corresponds to the case of $(S_{\rm{A}},S_{\rm{B}})=(1/2,1)$.
In our previous study, by using the modified spin-wave (MSW) method, we calculated the ferrimagnetic ground-state energy and obtained the phase boundary $\lambda_c=0.974$\cite{Hirose2017}. 

\begin{figure}[t]
\begin{center}
\includegraphics[width=.5\linewidth]{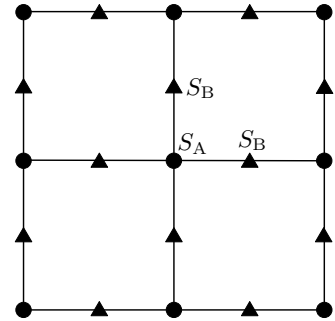}
\caption{Structure of a Lieb lattice, whose unit cell consists of three sites.
The closed circles and triangles represent A- and B-sublattice sites, respectively. 
The Marshall--Lieb--Mattis theorem ensures a ferrimagnetic ground state for ($S_{\rm{A}}$,$S_{\rm{B}}$)=($1/2,1$) and a singlet ground state for ($S_{\rm{A}}$,$S_{\rm{B}}$)=($1,1/2$). 
 }
\label{fig:2}
\end{center}
\end{figure}

First, in order to confirm the accuracy of the MSW results obtained in Ref.~3, such as $\lambda_c=0.974$, 
we calculate the ferrimagnetic ground-state energy and long-range order (LRO) parameters using the quantum Monte Carlo (QMC) method\cite{Todo} to compare them with the MSW results.

In Table I, we show the ground-state energy per site $e_g$ and the staggered magnetization $m_s$ for the $(S_{\rm{A}},S_{\rm{B}})=(1/2,1)$ system obtained using the QMC and MSW methods. 
For comparison, we show $e_g$ and $m_s$ for a spin-1/2 square-lattice antiferromagnet.\cite{Takahashi,Sauerwein}
In our QMC simulation, we calculate the structure factors $\mathcal{S}_{\rm{tot}}(N,T)=1/N\sum_{i,j}(-1)^{|i-j|}\langle \bm{S}_{i}\cdot\bm{S}_j\rangle$, 
where $N$ is the total number of sites and $\langle\cdots\rangle$ represents the thermal average at temperature $T$.
The staggered magnetization $m_s$ is obtained from $m_s^2=\lim_{N\to\infty}\lim_{T\to0}\mathcal{S}_{\rm{tot}}(N,T)/N$. 
For the definition of $L^2\equiv N/3$, we use three points $L=16, 24, 32$ to make linear extrapolation with respect to $L^{-1}$ $(L^{-3})$ for $\mathcal{S}_{\rm{tot}}/N$ (ground-state energy per site)\cite{Herbert}.
We set the temperature as $T=0.001$ in the unit of the exchange parameter in Eq.~(\ref{eq:1}), which can be regarded as absolute zero temperature in the system sizes we use.

In Table I, the numbers in parentheses next to the MSW results of $e_g$ represent the relative error with respect to the QMC results. 
Furthermore, the numbers in parentheses next to $m_s$ represent the ratio of $m_s$ to that of the classical vector model.
As for the staggered magnetizations $m_s$, in the case of the Lieb lattice, the ratios are approximately 92\%, while in the case of a square-lattice antiferromagnet, those are approximately 61\%.
We expect that a larger magnetization, i.e., a smaller spin deviation, makes spin-wave calculation more reliable.
The relative error of the ground-state energy on the Lieb lattice, $0.024\%$, is smaller than that of the square-lattice antiferromagnet, $0.21\%$, by one order of magnitude. 
The fact that the Lieb-lattice result is more accurate than the square-lattice result is consistent with the above-mentioned expectation from the staggered magnetizations.
Furthermore, we obtain the phase boundary between the macroscopically degenerated tetramer-dimer states and the ferrimagnetic ground state on a diamond-like-decorated square lattice\cite{Hirose2017}, $\lambda_c^{\rm{QMC}}=0.9739$ and $\lambda_c^{\rm{MSW}}=0.9743$, 
which show a good agreement within three digits after the decimal point.

\begin{table}[h]
\centering
\caption{Ground-state energy per site $e_g$ and staggered magnetization $m_s$, of the $(S_{\rm{A}},S_{\rm{B}})=(1/2,1)$ Lieb-lattice and spin-1/2 square-lattice antiferromagnets obtained using the MSW and QMC methods. 
The results of a spin-1/2 square-lattice antiferromagnet are cited from Refs.~7 and 8.}
\label{table:1}
\renewcommand{\arraystretch}{1.3}
  \begin{tabular}{@{\hspace{\tabcolsep}\extracolsep{\fill}}c|c|c|c} \hline
     & \text{lattice} & \text{QMC} & \text{MSW} \\ \hline 
$e_g$   & \text{Lieb} & $-0.820275(4)$ & $-0.82047$\;[$0.024\%$]\\ \cline{2-4}
      & \text{square} & $-0.6690(2)$ & $-0.67042$\;[$0.21\%$]  \\ \hline
$m_s$  & \text{Lieb} & $0.7650(3)$\;[$91.80\%$]  & $0.76557$\;[$91.87\%$] \\ \cline{2-4}
      & \text{square} & $0.307(6)$\;[$61.40\%$] & $0.3034$\;[$60.68\%$]\\ \hline
  \end{tabular}
\end{table}

Next, we consider the sublattice LRO parameters $m_{\rm{A}}^{(S_{\rm{A}})}$ for A-sublattice and $m_{\rm{B}}^{(S_{\rm{B}})}$ for B-sublattice.
In our previous MSW study, we found $\Delta m_{\rm{A}}^{(1/2)}=2\Delta m_{\rm{B}}^{(1)}$,
where $\Delta m_{\alpha}^{(S_{\alpha})}\equiv S_{\alpha}-m_{\alpha}^{(S_{\alpha})}$ ($\alpha={\rm A}, {\rm B}$) denotes the spin reduction, 
and pointed out that the factor 2 comes from the ratio of the sublattice sizes.\cite{Hirose2017}
In other words, there exists a scaling relation that the spin reduction of each sublattice is inversely proportional to the number of sublattice sites.\cite{Hirose2017}
Now, we turn to checking whether the scaling relation holds in the QMC calculations. 
By denoting the numbers of A- and B-sublattice sites as $N_{\rm A}(=N/3)$ and $N_{\rm B}(=2N/3)$,
we introduce structure factors $\mathcal{S}_{\rm{A}}=1/N_{\rm{A}}\sum_{i,j\in \rm{A}}\langle \bm{S}_{i}\cdot\bm{S}_{j}\rangle$ for A-sublattice
and $\mathcal{S}_{\rm{B}}=1/N_{\rm{B}}\sum_{i,j\in \rm{B}}\langle \bm{S}_{i}\cdot\bm{S}_{j}\rangle$ for B-sublattice. 
Thus, we can obtain the sublattice magnetizations from 
$\{m_{\alpha}^{(S_{\alpha})}\}^2=\lim_{N_{\alpha}\to\infty}\lim_{T\to0}\mathcal{S}_{\alpha}/N_{\alpha}$ ($\alpha={\rm A},\;{\rm B}$).
We present the calculated results in Table II and obtain $\Delta m_{\rm{A}}^{(1/2)}=0.1031(4)$ and $\Delta m_{\rm{B}}^{(1)}=0.0514(3)$, 
which indicates that the relation $\Delta m_{\rm{A}}^{(1/2)}=2\Delta m_{\rm{B}}^{(1)}$ holds in the range of the statistical error. 
Therefore, our QMC simulation confirms the scaling relation. 

\begin{table}[h]
\centering
\caption{Sublattice LRO parameters $m_{\rm{A}}^{(1/2)}$ and $m_{\rm{B}}^{(1)}$ for the $(S_{\rm{A}},S_{\rm{B}})=(1/2,1)$ system.}
\label{table:2}
\renewcommand{\arraystretch}{1.5}
\begin{tabular}{c|c|c} \hline
                 & \text{QMC} &  \text{MSW} \\ \hline
$m_{\rm{A}}^{(1/2)}$ & $0.3969(4)$ & $0.39835$ \\ \hline
$m_{\rm{B}}^{(1)}$ & $0.9486(3)$ &  $0.94918$ \\ 
\hline\end{tabular} 
\end{table}

Now, we turn to the mathematical proof of the scaling relation.
From the Marshall--Lieb--Mattis theorem, the total spin of the ground states of Eq.~(\ref{eq:1}) is $S_{\rm{tot}}=N_{\rm{B}}S_{\rm{B}}-N_{\rm{A}}S_{\rm{A}}$,
where $N_{\rm{A}}S_{\rm{A}}<N_{\rm{B}}S_{\rm{B}}$ is assumed.
If we apply a small uniform magnetic field in the $z$ direction, 
the magnetic field selects a unique ground state, in which the total $S^z$ is given by $S^z_{\rm{tot}}=N_{\rm{B}}S_{\rm{B}}-N_{\rm{A}}S_{\rm{A}}>0$.
By defining the spin reduction operators via $S_{i}^z=-(S_{\rm{A}}-\Delta S_{i}^z)$ for $i\in \rm{A}$ and $S_{j}^z=S_{\rm{B}}-\Delta S_{j}^z$ for $j\in \rm{B}$,
we have $\sum_{i\in \rm{A}}\Delta S_{i}^z=\sum_{j\in \rm{B}}\Delta S_{j}^z$ in the sector the unique ground state belongs to.
Then, we take the thermal average of this equation to get
\begin{align}
N_{\rm{A}}\Delta m_{\rm{A}}^{(S_{\rm{A}})}=N_{\rm{B}}\Delta m_{\rm{B}}^{(S_{\rm{B}})},
\label{eq:2}
\end{align}
which indicates that the spin reduction in each sublattice is inversely proportional to the sublattice size.
By setting $S_{\rm{A}}=1/2$ and $S_{\rm{B}}=1$ and using $N_{\rm{B}}/N_{\rm{A}}=2$ for the Lieb lattice, we obtain the expected formula $\Delta m_{\rm{A}}^{(1/2)}=2\Delta m_{\rm{B}}^{(1)}$.

We note that the ground state is uniquely determined to be a singlet state when $N_{\rm{A}}S_{\rm{A}}=N_{\rm{B}}S_{\rm{B}}$.
Because there exist no spin polarizations, $m_{\alpha}^{(S_{\alpha})}=S_{\alpha}-\Delta m_{\alpha}^{(S_{\alpha})}=0$ ($\alpha={\rm A},\;{\rm B}$), in the singlet ground state, we notice that Eq.~(\ref{eq:2}) is trivially satisfied and gives no additional information. 
We have to apply a staggered magnetic field instead of a uniform field or look into the LROs to get meaningful sublattice magnetizations.
In order to study this issue, we consider another case of $(S_{\rm{A}},S_{\rm{B}})=(1,1/2)$, which gives $N_{\rm{A}}S_{\rm{A}}=N_{\rm{B}}S_{\rm{B}}$, and carry out the MSW and QMC calculations.
In Table III, we show the ground-state energy $e_g$, staggered magnetization $m_s$, and 
sublattice LRO parameters $m_{\rm{A}}^{(1)}$ and $m_{\rm{B}}^{(1/2)}$.
First, by considering the relative error of $e_g$ and ratio of $m_s$ to that of the classical vector model, 
we find that the accuracy of the case $(S_{\rm{A}},S_{\rm{B}})=(1,1/2)$ is comparable to that of the spin-1/2 square-lattice antiferromagnet.
We find from Table III that 
\begin{equation}
   \frac{\Delta m_{\rm{A}}^{(1)}}{\Delta m_{\rm{B}}^{(1/2)}}=\begin{cases}
								\frac{0.3782(6)}{0.1890(3)}=2.001(6)&\;\text{(QMC)}\\
								\frac{0.40917}{0.20459}=2&\;\text{(MSW)}
										\end{cases}
\label{eq:3},																
\end{equation}
and thus the scaling relation $\Delta m_{\rm{A}}^{(1)}=2\Delta m_{\rm{B}}^{(1/2)}$ is confirmed.
In the MSW theory, the scaling relation holds exactly in analytical expressions.
The result in Eq.~(\ref{eq:3}) indicates that the above-mentioned proof for spontaneous magnetizations is insufficient, and it is desired to find a proof with respect to the LROs and/or to introduce a staggered magnetic field. 

\begin{table}[h]
\centering
\caption{Calculated results for the $(S_{\rm{A}},S_{\rm{B}})=(1,1/2)$ system with the singlet ground state.}
\label{table:3pp}
\renewcommand{\arraystretch}{1.5}
\begin{tabular}{c|c|c} \hline
   &  \text{QMC}  &    \text{MSW} \\ \hline
$e_g$ &                       $-0.89581(1)$           & $-0.89902\;[0.36\%]$  \\ \hline 
$m_s$ &                      $0.4145(4)\;[62.18\%]$ & $0.39389\;[59.08\%]$ \\  \hline
$m_{\rm{A}}^{(1)}$ &     $0.6218(6)$ & $0.59083$ \\  \hline
$m_{\rm{B}}^{(1/2)}$ &  $0.3110(3)$ & $0.29541$ \\  \hline
\end{tabular} 
\end{table}

For the ground states of the Hamiltonian in Eq.~(\ref{eq:1}), the Marshall--Lieb--Mattis theorem gives
\begin{align}
\bm{S}_{\rm{tot}}^2=(N_{\rm{A}}S_{\rm{A}}-N_{\rm{B}}S_{\rm{B}})(N_{\rm{A}}S_{\rm{A}}-N_{\rm{B}}S_{\rm{B}}+1)
\label{eq:10}.
\end{align}
However, if a small staggered magnetic field is applied, then the total spin is not conserved.
In this case, we have to deal with the following ground-state expectation value,
\begin{align}
\langle\bm{S}_{\rm{tot}}^2\rangle_g=\sum_{i,j}\langle\bm{S}_i\cdot\bm{S}_j\rangle_g\equiv aN^2+\mathcal{O}(N)
\label{eq:11}.
\end{align}
In the MSW theory, it is known that the Dyson--Maleev transformation is implicitly accompanied with a symmetry-breaking field and breaks the conservation of the total spin.
It is very instructive to examine the mathematical structure behind the scaling relation in the MSW theory.
Therefore, within the MSW theory, we intend to calculate
\begin{align}
   a=\lim_{N\to\infty}\frac{1}{N^2}\sum_{i,j}\langle\bm{S}_i\cdot\bm{S}_j\rangle_g
\label{eq:12}.
\end{align}
This coefficient relates to the LROs. The sublattice magnetizations, $m_{\rm{A}}$ and $m_{\rm{B}}$, are defined by
\begin{equation}
   \lim_{|i-j|\to\infty}\langle\bm{S}_i\cdot\bm{S}_j\rangle_g=\begin{cases}
										m_{\rm{A}}^2 &\;\text{$(i,j \in \rm{A})$}\\
										m_{\rm{B}}^2 &\;\text{$(i,j \in \rm{B})$}
										\end{cases}	
\label{eq:13}.																			
\end{equation}
In addition, the MSW calculation yields
\begin{align}
\lim_{|i-j|\to\infty}\langle\bm{S}_i\cdot\bm{S}_j\rangle_g=-m_{\rm{A}}m_{\rm{B}}\;\;\;\text{($i\in\rm{A}$, $j\in\rm{B}$ or v.v.)},
\label{eq:14}
\end{align}
which is shown in our previous paper\cite{Hirose2017}. 
By evaluating the right-hand side of Eq. (\ref{eq:12}), we obtain
\begin{eqnarray}
   a&&\hspace{-6mm}=\frac{N_{\rm{A}}^2m_{\rm{A}}^2+N_{\rm{B}}^2m_{\rm{B}}^2-2N_{\rm{A}}N_{\rm{B}}m_{\rm{A}}m_{\rm{B}}}{N^2}
   \nonumber\\
   &&\hspace{-6mm}={\left(\frac{N_{\rm{A}}m_{\rm{A}}-N_{\rm{B}}m_{\rm{B}}}{N_{\rm{A}}+N_{\rm{B}}}\right)}^2
\label{eq:15},
\end{eqnarray}
because of large distances between sites $i$ and $j$ in almost all terms in the summation.
In the MSW theory, we have the scaling relation of the spin reductions, 
\begin{align}
   N_{\rm{A}}(S_{\rm{A}}-m_{\rm{A}})=N_{\rm{B}}(S_{\rm{B}}-m_{\rm{B}})
\label{eq:16}.
\end{align}
By substituting Eq. (\ref{eq:16}) into Eq. (\ref{eq:15}), we get
\begin{align}
   a={\left(\frac{N_{\rm{A}}S_{\rm{A}}-N_{\rm{B}}S_{\rm{B}}}{N_{\rm{A}}+N_{\rm{B}}}\right)}^2
\label{eq:17},
\end{align}
which is the same as that obtained using the Marshall--Lieb--Mattis theorem. 

By reviewing the discussion above based on the MSW theory, we notice that Eqs. (\ref{eq:14}) and (\ref{eq:17}) are sufficient conditions for the scaling relation (\ref{eq:16}). In association with Eq. (\ref{eq:17}), we expect 
\begin{eqnarray}
   \langle\bm{S}_{\rm{tot}}^2\rangle_g&&\hspace{-6mm}=(N_{\rm{A}}S_{\rm{A}}-N_{\rm{B}}S_{\rm{B}})^2
   \nonumber\\
   &&\hspace{0mm}+\text{[field-dependent $\mathcal{O}(N)$ terms]},
\label{eq:18}
\end{eqnarray}
even if a symmetry-breaking field prevents the application of the Marshall--Lieb--Mattis theorem. In a system with a ferrimagnetic ground state, both uniform and staggered magnetic fields can lead to the appearance of spontaneous symmetry breaking, and it is naturally expected that the choice of a symmetry-breaking field affects only the $\mathcal{O}(N)$ part. For the case of $N_{\rm{A}}S_{\rm{A}}-N_{\rm{B}}S_{\rm{B}}=0$, where the Marshall--Lieb--Mattis theorem predicts a singlet ground state, an infinitesimal staggered field leads to $\langle\bm{S}_{\rm{tot}}^2\rangle\neq0$ and makes a spontaneous staggered magnetization appear. A nonzero value of $\langle\bm{S}_{\rm{tot}}^2\rangle$ is expected to come from the $\mathcal{O}(N)$ part. As for the relation in Eq. (\ref{eq:14}), it might be trivial for a classical vector counterpart of the Heisenberg model, but the mathematical proof for the present quantum system is an open issue. If Eq. (\ref{eq:14}) was proved, we would arrive at the mathematical proof of the scaling relation concerned with the LROs.
Our QMC simulation shows that Eq. (\ref{eq:14}) holds within the statistical error.

In summary, we calculated the ground-state energy and LRO parameters of mixed spin-1 and spin-1/2 Lieb-lattice antiferromagnets using the QMC method and compared the obtained results with the MSW results. 
Furthermore, we showed the scaling relation for the sublattice spin-reductions, which can be proved by applying a uniform field to cause spontaneous symmetry breaking. 
However, we found that the scaling relation holds even in the singlet ground state, where uniform fields do not give a spontaneous symmetry breaking. 
Therefore, in the MSW theory, we discussed the scaling relation for the LRO parameters and found that, if Eq. (\ref{eq:14}) was proved, we would arrive at the mathematical proof of the scaling relation for the LRO parameters including the singlet ground state.

We acknowledge Professor K. Hida for fruitful discussions, which brought the total spin of the MSW ground state to our attention, and Professor A. Oguchi for helpful discussions including our previous studies. 
The authors thank the Supercomputer Center, the Institute for Solid State Physics, the University of Tokyo for the use of the facilities.
This work was partly supported by JSPS KAKENHI Grant Numbers JP17J05190 and JP17K05519.


\begin{thebibliography}{99} 
\bibitem{Heidelberg} Introduction to Frustrated Magnetism, ed. C. Lacroix, P. Mendels, and F. Mila (Springer, Heidelberg, 2011) Springer Series in Solid-State Sciences, Vol. 164.
\bibitem{Hirose2016}Y. Hirose, A. Oguchi, and Y. Fukumoto, J. Phys. Soc. Jpn. \textbf{85}, 094002 (2016).
\bibitem{Hirose2017} Y. Hirose, A. Oguchi, and Y. Fukumoto, J. Phys. Soc. Jpn. \textbf{86}, 014002 (2017).
\bibitem{QDM} D. S. Rokhsar and S. A. Kivelson, Phys. Rev. Lett. \textbf{61}, 2376 (1988).
\bibitem{Morita} K. Morita and N. Shibata, J. Phys. Soc. Jpn. \textbf{85}, 033705 (2016).
\bibitem{Todo} S. Todo and K. Kato, Phys. Rev. Lett. \textbf{87}, 047203 (2001).
\bibitem{Takahashi} M. Takahashi, Phys. Rev. B \textbf{40}, 2494 (1989).
\bibitem{Sauerwein} R. A. Sauerwein and M. J. de Oliveira, Phys. Rev. B \textbf{49}, 5983 (1994).
\bibitem{Herbert} H. Neuberger and T. Ziman, Phys. Rev. B \textbf{39}, 2608 (1989).
\end{thebibliography}
\end{document}